\documentclass[journal=ancac3,manuscript=article]{achemso}

\usepackage{chemformula} % Formula subscripts using \ch{}
\usepackage[T1]{fontenc} % Use modern font encodings

\newcommand{\yz}[1]{\textcolor{blue}{#1}}
\usepackage{ulem}
\author{Xiaoyan Wu}
\affiliation{Shenzhen JL Computational Science and Applied Research Institute, Longhua District, Shenzhen 518110, China}

\author{Tammo van der Heide}
\affiliation{Bremen Center for Computational Materials Science, University of Bremen, 28359 Bremen, Germany}

\author{Shizheng Wen}
\affiliation{Jiangsu Province Key Laboratory of Modern Measurement Technology and Intelligent Systems, School of Physics and Electronic Electrical Engineering, Huaiyin Normal University, 223300, Huaian, China}

\author{Thomas Frauenheim}
\affiliation{Beijing Computational Science Research Center, Haidian District, Beijing 100193, China}
\alsoaffiliation{Bremen Center for Computational Materials Science, University of Bremen, 28359 Bremen, Germany}
\alsoaffiliation{Shenzhen JL Computational Science and Applied Research Institute, Longhua District, Shenzhen 518110, China}

\author{Sergei Tretiak}
\affiliation{Theoretical Division, Los Alamos National Laboratory, Los Alamos, New Mexico 87545, United States }
\alsoaffiliation{Center of Integrated Nanotechnlogies, Los Alamos National Laboratory, Los Alamos, New Mexico, 87545, United States}

\author{ChiYung Yam}
\affiliation{Shenzhen Institute for Advanced Study, University of Electronic Science and Technology of China, Shenzhen, 518000, China }
\email{yamcy@uestc.edu.cn}

\author{Yu Zhang}
\affiliation{Theoretical Division, Los Alamos National Laboratory, Los Alamos, New Mexico 87545, United States }
\email{zhy@lanl.gov}

\title[An \textsf{achemso} demo]
  {Molecular Dynamics Study of Plasmon-Mediated Chemical Transformations} %Activation of CO}


\begin{document}

\begin{tocentry}
% A TOC is needed!!.
% The toc is not good enough. 
%\includegraphics[width=0.99\textwidth]{TOC.pdf}
\centering
\includegraphics{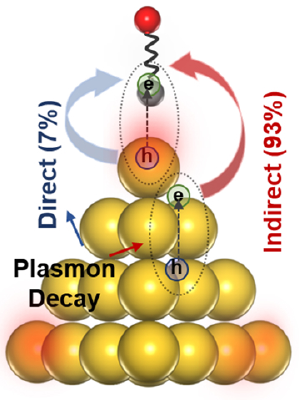}
\end{tocentry}

\begin{abstract}
Heterogeneous catalysis of adsorbates on metallic surfaces mediated by plasmon has potential high photoelectric conversion efficiency and controllable reaction selectivity. Theoretical modeling of dynamical reaction processes provides in-depth analyses complementing experimental investigations. Especially for plasmon-mediated chemical transformations, light absorption, photoelectric conversion, electron-electron scattering, and electron-phonon coupling occur simultaneously at different timescales, rendering it very challenging to delineate the complex interplay of different factors. In this work, a trajectory surface hopping non-adiabatic molecular dynamics method is used to investigate the dynamics of plasmon excitation in an Au$_{20}$-CO system, including hot carrier generation, plasmon energy relaxation, and CO activation induced by electron-vibration coupling. The electronic properties indicate that when Au$_{20}$-CO is excited, a partial charge transfer takes place from Au$_{20}$ to CO. On the other hand, the dynamical simulations show that hot carriers generated after plasmon excitation transfer back and forth between Au$_{20}$ and CO. Meanwhile, the C-O stretching mode is activated due to the non-adiabatic couplings. The efficiency of plasmon-mediated transformation ($\sim$40\%) is obtained based on the ensemble average of these quantities. Our simulations provide important dynamical and atomistic insights into plasmon-mediated chemical transformation from the perspective of non-adiabatic simulations. 
\end{abstract}

\section{Introduction}
Plasmon-mediated chemical transformations, which utilize plasmonic nanostructures as catalysts, is an emerging technology that has attracted extensive attentions~\cite{2011natchem,Linic2016natcom,RN3,Halas2021pnas,2020japreview,MRS2020bulletin,clavero2014natphotonics,Kim2019angew,zhao2019jpcc,Kumari2018acsnano,shaik2019nl,am202008145,jpca.1c05129,jpcc.1c07575,natcom2017Narang}. Here, energy of light is collected by the collective oscillations of electrons, resulting in a localized plasmon resonance (LSP) which has scattering cross section much larger than relevant geometric sizes~\cite{plasmon_anrev,jp026731y,cr100313v}. After being excited, LSP decays simultaneously through radiative and nonradiative pathways. Usually, the majority of the energy stored in the plasmonic field is dissipated through nonradiative decay, resulting in the formation of energetic (hot) electrons (HEs) and holes~\cite{RN1}. Surface scattering, electron-phonon coupling, and electron-electron scattering are the major mechanisms that determine the distribution of these carriers~\cite{chang2019electronic,Bernardi2015nc,acsnano5b06199,natcomm2014Atwater}. The energies of HEs (and their corresponding holes) are then redistributed via electron-electron scattering, released to phonon modes via electron-phonon interactions, and ultimately dissipated into environment via thermal conduction~\cite{acsph2017govorov,zhang2021jpca}.

Thanks to their tunable optical properties, plasmons have been extensively explored in the context of driving chemical transformations via different mechanisms. So far, four major mechanisms have been identified in these reactions~\cite{Kim2019angew}, including 1) plasmon-enhanced intramolecular excitations (or near-field effect), where the plasmon resonance overlaps with the electronic transition energies~\cite{Kim2018science,Terefer2017jpcc,Fiona2022JMCC,Christopher2019acseng}, 2) indirect HE transfer from metal nanoparticles (NPs) to the adsorbed molecules, where HE generation within the nanostructures follows by a charge transfer process to the adsorbed molecules~\cite{wu2020jacs,Preahant2018nl,RN6}, 3) direct charge transfer mechanism, where the electrons are directly excited from the valence band of plasmonic materials to the virtual molecular orbital (MO) of the adsorbates~\cite{Lian2015science,Oleg2012jacs,Oleg2014jacs,RN13}, and 4) thermal activation (or local heating effect) due to the HE relaxation~\cite{heating2018jacs,Katherine2018jpcc,Yonatan2021ACSphtonics}. In particular, HE-mediated reactions have higher tunability compared to conventional temperature-driven catalysis because the energy can be selectively deposited into particular reaction coordinates in the former process. 

Hence, HE-mediated chemical transformation of adsorbates on the surface of nanoparticles induced by the unique light-matter interactions of plasmonic NPs has attracted significant attention in recent decades. For instance, plasmonic excitations of Ag and Au NPs have shown to activate carbonyl hydrogenation, dissociation of H$_{2}$ and reduction of nitroaromatics, etc.~\cite{RN6,RN7,RN8,RN9,RN10,RN11}. In addition, for HE-mediated processes, recent reports have suggested that interactions with semiconductors or adsorbates enhance the reaction efficiency through interfacial states~\cite{RN26,cid2019scfadv}. These interfacial states provide an additional energy dissipation pathway for the plasmons or serve as transient reservoirs of these hot carriers.\cite{RN3,Linic2016natcom,RN26} Other studies have investigated the tunability of energy distribution of hot carriers to enhance quantum yield and also to realize controllability of reaction pathways. For example, Manjavacas and co-workers found that small-size NPs and longer carrier lifetimes result in higher hot carrier energies but lower hot-carrier production rates, and vice versa.\cite{RN1} Numerous studies have reported ways of improving HE production in plasmonic nanomaterials, including the use of small NPs with large surface-to-volume ratio, designing plasmonic materials with a longer carrier mean free path, constructing hybrid nanostructures with plasmon resonances in the red and infrared regions, and designing NPs with extended plasmonic hot spots~\cite{acsph2017govorov,acsph6b01048,natcomm2014Atwater,acsnano5b06199}. These findings have inspired further interests in the field of plasmonic catalysis as it offers opportunities to develop new and selective catalytic processes that were previously inaccessible.

In addition to HE-mediated reactions, heating effect due to the HE relaxation may activate chemical reactions as well. Recent debates on the thermal impact underpin the necessity of atomistic and dynamical insights via simulations of the HE generation, transfer, and relaxation processes on equal footing. This demands for special theoretical tools such as our recent development of NEXMD-DFTB method~\cite{wu2022nonadiabatic}. Dynamic simulations of plasmon-enhanced catalysis were previously performed using different theoretical methods. For example, Meng and co-workers have investigated H$_{2}$O splitting catalyzed by Au$_{20}$ cluster using real-time time-dependent density functional theory (RT-TDDFT) and Ehrenfest dynamics.\cite{RN14} Their results show significant energy and spatial overlap between oscillating electron density within the Au cluster and MOs of H$_{2}$O. Our previous work demonstrated that HEs transfer to the antibonding state of the H$_{2}$ molecule from the metal NP upon photoexcitation, a process that introduces a repulsive potential and drives the cleavage of the H-H bond.\cite{RN15} These calculations assume high-intensity photoexcitation, which depends on the strength of the external field. Recently, we also proposed a mechanism for H$_2$ dissociation, which suggests that the charge transition from HE states within the metal NP to charge transfer (CT) excited states is the main channel to trigger H$_2$ reaction.\cite{wu2020jacs} Importantly, this study only involved one electron (one-hole) or mono-electronic excitation process, which facilitates the understanding of the basic principle of such reactions. Furthermore, this investigation explained the energy and electron transfer in terms of reaction coordinates, which elucidates the details of reaction pathway. However, the simple model system in this study does not support plasmon excitation, and therefore precludes the description of processes including plasmon decay into hot carriers, electron transfer to adsorbed H$_2$, and hot carriers decay through electron-phonon interaction.

In this work, an Au$_{20}$ cluster is chosen, which supports plasmon-like excitation with a symmetric geometry and high stability~\cite{RN17}. A CO molecule is adsorbed on the Au$_{20}$ which gives rise to CT states~\cite{RN18}. 
For this system, the coupling strength between Au$_{20}$ and CO is in the intermediate regime, so both direct and indirect mechanisms can be studied simultaneously, together with their competition with HE relaxation. Through molecular dynamics simulations, it is find that the C-O stretching vibration is excited by HE transfer via CT excited states. These states are excited via either direct or indirect HE transfer processes. Moreover, our results reveal that both direct and indirect processes are significantly faster than the energy relaxation process in Au$_{20}$, which demonstrates that the HE transfer is the dominating process that governs the chemical transformation of the CO molecule.

\section{Methods}\label{sec_method}
\textbf{Density Functional Tight-Binding Theory (DFTB) Calculations}. Geometry optimizations and ground-state properties are carried out with the DFTB+ code.\cite{RN19} The auorg-1-1 parameter set, designed to describe optical excitations of organic molecules on gold nanoclusters, are employed for all computations in this work~\cite{RN20}. To obtain a correct desorption reaction coordinate, the DFTB parameters associated with the repulsive potential between Au and C (O) atoms are modified (more details in the \textbf{Results and Discussions} section). The electron occupation for each Kohn–Sham (KS) orbital is estimated by Mulliken charge analysis.\cite{RN19} Only valence electrons are considered in DFTB method, specifically, nine, six and four valence electrons for Au, O and C atoms are included, respectively. Therefore, the Au$_{20}-$CO system has 188 KS orbitals and 115 of them are occupied.

\textbf{Time-Dependent Density Functional Tight-Binding Theory (TDDFTB) Calculations}. Linear response time-dependent DFTB (LR-TDDFTB) calculations within the random phase approximation (RPA) formalism are performed to obtain excitation energies, oscillator strengths, and coefficients associated with the contribution of a given KS orbital transition of each excitation.\cite{RN21,kranz2017time} The optical spectra are broadened using Gaussian function with a half-width of 0.1~eV. The adiabatic excitation energies are obtained via an eigenvalue equation, $\Lambda \, F_{I} = \Omega_I^2 \, F_{I}$, where $\Omega_I$ denotes the energy of $I^{th}$ excitation. $\Lambda$ is the RPA matrix with its elements in MO representation given by $\Lambda_{ia,jb} = \omega_{ia}^2\delta_{ij} \delta_{ab} + 4\sqrt{\omega_{ia}}  K_{ia,jb} \sqrt{\omega_{jb}}$,
$\omega_{ia}\,= \epsilon_a \, - \epsilon_i$ and $\epsilon_{i/a}$ is the KS orbital energy. $\{i, j \, \ldots\}$ and $\{a, b , \cdots\}$ denote the occupied and virtual KS orbitals, respectively. $K$ is the exchange-correlation kernel. Within the Mulliken approximation and LR-TDDFTB formalism, the kernel is simplified by employing the $\gamma$ approximation~\cite{RN21}, $K_{ia,jb} = \sum_{\mu \, \nu} \, q_\mu^{ia} \, \gamma_{\mu \, \nu} \, q_\nu^{jb}$. Here $q_\mu^{ia}$ denotes the Mulliken transition charges $q_\mu^{ia} \, = \frac{1}{2}\sum_\nu \, (c_{\mu \, i} \, c_{\nu \, a} \, S_{\mu\nu} \, + c_{\nu \, i} \, c_{\mu \, a} \, S_{\nu\mu})$, $S$ is the overlap matrix of atomic orbitals, $\{\mu,\nu, \cdots\}$ denote atomic orbitals, and $\gamma$ includes the Coulomb and gradient-corrected exchange-correlation kernels. The element $F_{ia}$ denotes the contribution coefficient for KS orbital transition from an occupied orbital $i$ to an virtual orbital $a$. Hence, for the Au$_{20}-$CO system, 8395 possible transitions are included for each excited state. Because plasmon excitation usually corresponds to higher energy excited state, hundreds of excited states are generally required. To reduce the computational cost, only occupied-virtual MO pairs with energy differences smaller than 3.85~eV are included in the excited state calculations. 

\textbf{Trajectory Surface Hopping (TSH) non-adiabatic Molecular Dynamics (NAMD) Simulations}. 
The TSH driver within the LR-TDDFTB framework is used to simulate the NAMD of Au$_{20}-$CO initiated by plasmon excitation. This is recently implemented by combining the open-source software packages DFTB+ and NEXMD.~\cite{wu2022nonadiabatic} The adiabatic electronic structures, including ground, excited state energies, gradients, excited states transition densities, and non-adiabatic couplings (NACs) between excited states, are described at the TDDFTB level. The non-adiabatic effects originating from coupled motions of electrons and nuclei are treated by the TSH algorithm~\cite{wu2022nonadiabatic,NEXMD_plasmonics,sergei2020chemrev}. First, the system is optimized on its singlet ground state at the DFTB level. This is followed by a 90~ps Born-Oppenheimer ground state MD simulation with a time step $\Delta t=0.1$~fs. Here, a Langevin thermostat with a damping rate of 20~ps$^{-1}$ is used to keep the temperature fluctuating around 300~K.~\cite{RN22} After a 10~ps equilibration period, 300 snapshots of initial geometries together with nuclear velocities are sampled from the ground state trajectory with 166~fs intervals, which are used to calculate the average absorption spectrum and serve as starting points for subsequent excited state dynamics. To obtain the average absorption spectrum, vertical excitation energies and oscillator strengths of all snapshots are computed using the LR-TDDFTB.~\cite{RN20} For NAMD simulations, all excited state trajectories are prepared by plasmon excitation (i.e., 2.7~eV) according to the absorption spectrum and are propagated for 1~ps. The timesteps for nuclear (classical) and electronic (quantum) equations of motions are set to 0.4~fs and 0.1~fs, respectively. 160 excited states are taken into account ($N_{st} =160$) in our simulations to ensure the inclusion of plasmon excitation and allow for possible transitions to higher excited states. As described in previous work, the instantaneous decoherence corrections (IDC) and trivial unavoided crossings methods are adopted to improve the accuracy of calculations~\cite{RN22}. The quantum timestep is further refined by a factor of 10 when potential trivial crossings are detected.

\begin{figure}[!htb]
   \centering
   \includegraphics[width=0.5\textwidth]{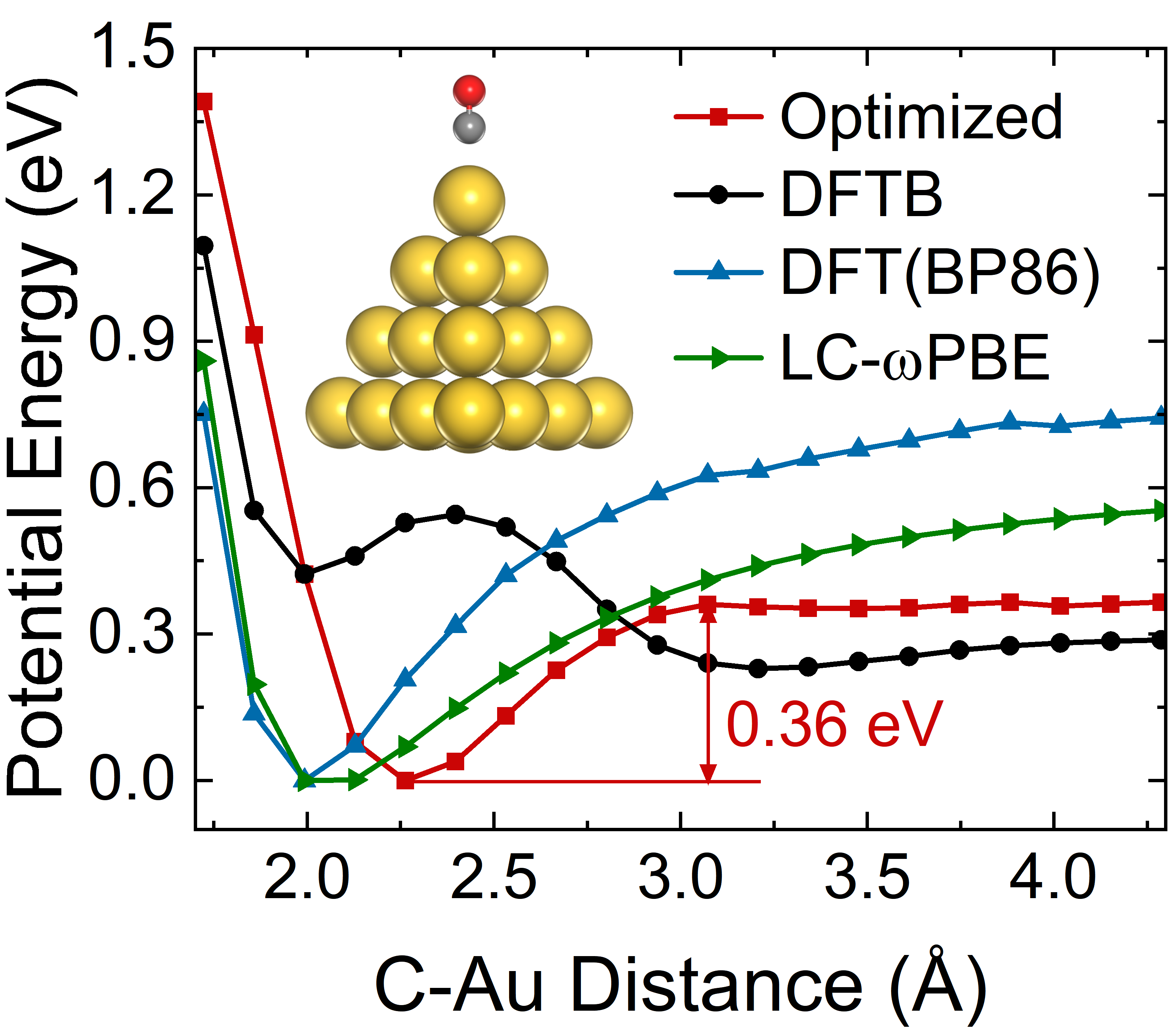}
   \caption{
   Ground state PES along the reaction coordinate of CO on Au$_{20}$. CO molecule is placed directly above the apex site of Au$_{20}$ and the distance with respect to Au$_{20}$ is varied along the vertical direction. Inset shows the ground state equilibrium structure of Au$_{20}-$CO, where yellow, red, and brown spheres represent gold, oxygen, and carbon atoms, respectively.}
   \label{Fig.1}
\end{figure} 

\section{Results and Discussions}
\subsection{Geometry and Electronic Properties of Au$_{20}-$CO}
Different conformations of adsorbates on the NP surface strongly influence the electronic coupling, ultimately determine the mechanism of hot carrier transfer and plasmon decay.\cite{persson1993polarizability, RN23, Linic2016natcom} The inset of Figure~\ref{Fig.1} shows the ground state equilibrium geometry of Au$_{20}-$CO optimized by DFTB method.\cite{RN20} This structure with CO molecule adsorbed at the apex site of the Au$_{20}$ cluster is consistent with previous findings.\cite{RN24} To further confirm this adsorption configuration is favorable during the dynamic process, ground state potential energy surface (PES) along the adsorption reaction coordinate (Figure~\ref{Fig.1}) is calculated. As compared with DFT (BP86~\cite{becke1988density}) result, it can be clearly seen that the auorg-1-1 parameter set~\cite{RN20} significantly underestimates the energy barrier of CO desorption. Therefore, in order to obtain a correct topology of PES, we reoptimize the repulsive potential between atom C (O) and Au in the auorg-1-1 parameter set.~\cite{RN20, DFTBpara} As shown in Figure~\ref{Fig.1}, the potential profile obtained using the optimized parameter is consistent with that of DFT (BP86~\cite{becke1988density} and LC-$\omega$PBE~\cite{vydrov2006assessment}) results. The corresponding adsorption barrier of the CO molecule on the apex site of Au$_{20}$ is about 0.36~eV, which is characteristic of chemisorption.

\begin{figure}
    \centering
    \includegraphics[width=0.95\textwidth]{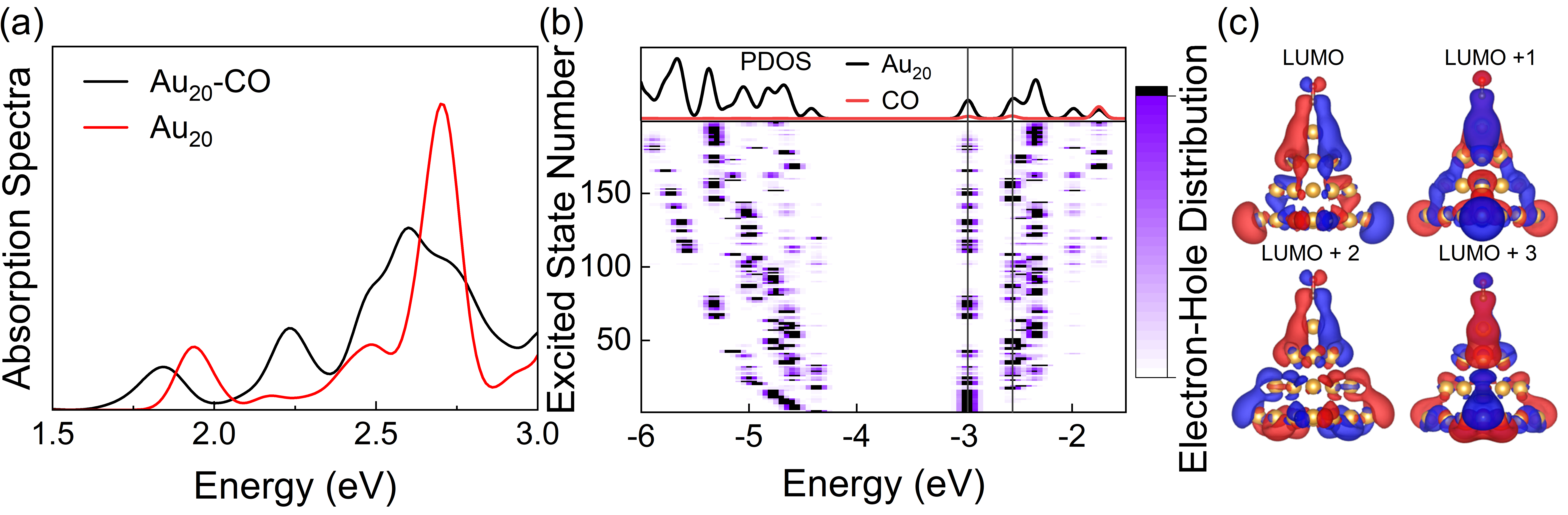}
    \caption{
    (a) Absorption spectra of Au$_{20}$-CO and bare Au$_{20}$ cluster obtained with LR-TDDFTB calculations. The spectra are broadened using a Gaussian function with a half-width of 0.1~eV. (b) Top panel: Projected density of states (PDOSs) of Au$_{20}$-CO. Bottom panel: Electron-hole distribution of excited states. Black dash lines indicate the position of the delocalized KS states. (c) Spatial distributions of delocalized KS MOs involved in the CT excited states. Calculations are performed based on ground-state geometry.}
  \label{Fig.2}
\end{figure}

With the optimized DFTB parameters, we first examine the static electronic properties of plasmon excitation of Au$_{20}$-CO. The optical absorption spectrum shows a dominant peak at 2.71~eV for a bare Au$_{20}$ cluster (red curve in Figure~\ref{Fig.2}(a)), showing a good agreement with the value of 2.78~eV in literature.\cite{RN17} The frequency-domain LR formalism, which describes excitations as weighted combinations of KS transitions between occupied and virtual MOs, permits an analysis of the collective character of plasmon excitation. As shown in Table S1, the excitation energy of 2.71~eV composes of multiple KS transitions with comparatively similar weights. 
The energy distributions of holes and electrons resulting from these multi-configurational transitions contain only a few peaks and have different shapes owing to the quantum confinement effect and discretized energy levels (Figure S1). For Au$_{20}$-CO, the adsorption of CO induces a broadening and slightly red shift of the major absorption peak (black curve in Figure~\ref{Fig.2}(a)), which are consistent with previous studies~.\cite{kreibig2013optical, persson1993polarizability} New peaks also emerge in the lower energy range of the absorption spectrum of Au$_{20}$-CO, which are absent in the absorption spectra of bare Au$_{20}$ and CO molecule. This signifies that the presence of CT excited states upon CO adsorption. Note that CT excited states defined here refer to excited states for which at least 0.1e is transferred from the Au$_{20}$ cluster to the CO molecule upon excitation (Table S4).
In combination with the analysis of transition component of the first excited state ($S_1$, Table S3) and the spatial distributions of MOs (Figure~\ref{Fig.2}(c)), we confirm that the KS transitions involved in the CT excited states are associated with KS MOs which are delocalized over both the Au$_{20}$ cluster and the CO molecule. As shown in Figure~\ref{Fig.2}(c), these virtual MOs (LUMO--LUMO+3) of interest are delocalized over both CO and Au$_{20}$ and show a significant $\pi^*$ character on the CO molecule. Furthermore, the electron-hole distribution of excited states ($S_1$- $S_{200}$) at the ground state geometry are depicted in Figure~\ref{Fig.2}(b). It is observed that a majority of excited states, including the plasmon excitation state ($S_{116}$, Figure~S1), involves these delocalized KS MOs. (The details of calculating the electron-hole distribution are provided in the Supporting Information.) It is worth noting that the charge transfer from Au$_{20}$ to CO by excitation of CT states can lead to the coupling between electrons and CO vibration.~\cite{JCP1991}

It is well-known that TDDFT calculations with local and semilocal functionals lead to spurious low-lying CT states. The parameterization in DFTB is however generally based on PBE semilocal functional which inherits this error in the description of CT states. Figure S2 shows the absorption spectra of Au$_{20}$ calculated with TDDFTB, TDDFT(PBE),~\cite{perdew1996generalized} and TDDFT(LC-$\omega$PBE).~\cite{vydrov2006assessment} As expected, the TDDFTB and TDDFT(PBE) spectra are densely populated with low-energy excited states as compared with that of long-range corrected LC-$\omega$PBE calculation. However, the sparse manifold of excited states calculated with LC-$\omega$PBE functional does not fit the characteristics of plasmon excitations. On the other hand, DFTB$\slash$TDDFTB with GGA functional are more suitable for qualitative description of the dense manifold of excited states in plasmon nanoparticles and the competition between HE relaxation and HE transfer-induced chemical reactions. Therefore, in this work, we study the plasmon-like excitation dynamics of the Au$_{20}$ systems at the level of TDDFTB. Nevertheless, for a more accurate description of excited state properties, long-range corrected DFTB can be applied which exhibits similar accuracy as range-separated DFT methods at significantly reduced computational cost.~\cite{koppen2012optical, silverstein2010assessment}

{
To summarize, the excited state properties of the Au$_{20}$-CO system indicate the existence of two qualitatively different manifolds of states. One manifold of excited states is dominated by the electronic excitation confined in the Au$_{20}$ cluster alone, which can be likened to the HE state. The other manifold involves the excitations to the delocalized hybrid Au$_{20}-CO$ MOs, which is likened to CT states~\cite{wu2020jacs}. Since HE and CT states can cross, indirect HE transfer from Au$_20$ cluster to CO molecule can occur via the non-adiabatic transitions between HE and CT states~\cite{wu2020jacs}. Hence, both direct and indirect HEs transfer may contribute to the relaxation of the plasmon excitation and activation of CO vibrations. 
At the heart of both direct and indirect HEs transfer mechanisms is the coupling between CT states and vibrational states of Au$_{20}$-CO complex, which can lead HEs to dispose a portion of their energy into the vibrational motion of CO during the relaxation process. To verify this scenario and to clarify the dynamical details, including the direct/indirect HE transfer, the non-adiabatic relaxation, and the competition between the HE transfer and energy relaxation, we next perform direct NAMD simulations of plasmon excitation. 
}

\begin{figure}[htb]
   \centering
   \includegraphics[width=0.95\textwidth]{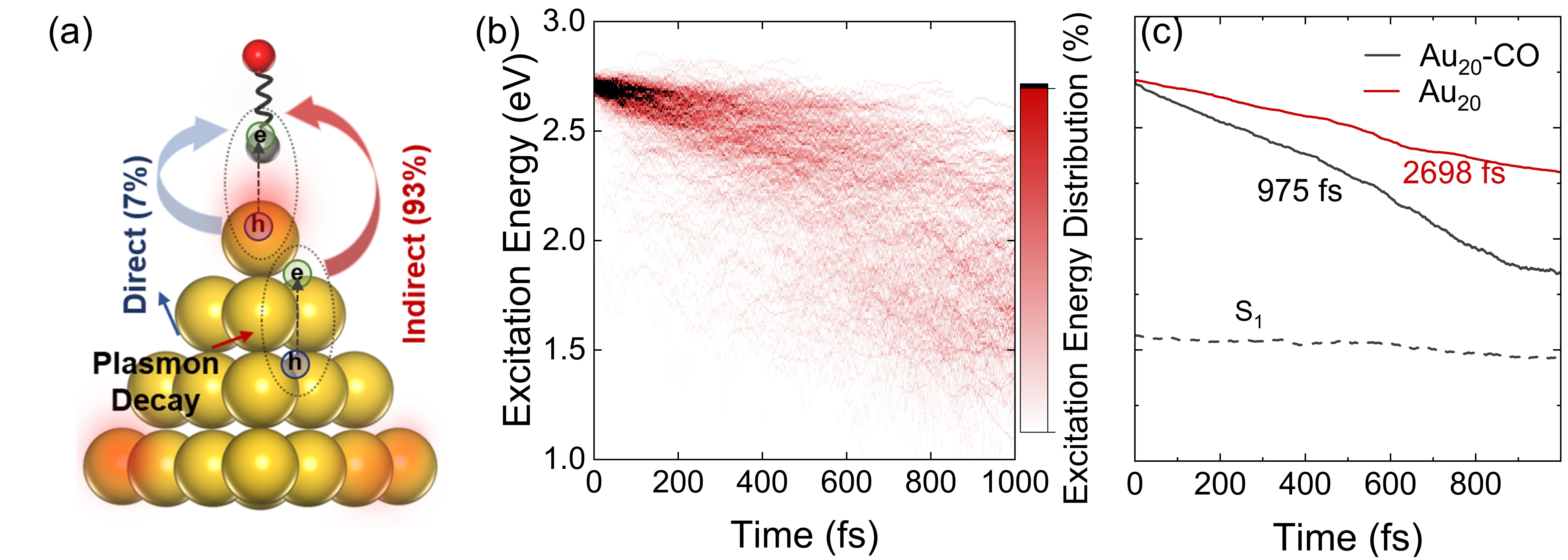}
   \caption{
   (a) Schematic of HE transfer process in Au$_{20}$-CO. (b) Relaxation time-energy 2D map for Au$_{20}$-CO. (c) Evolution of the average electronic energies of the Au$_{20}$-CO and bare Au$_{20}$.
   }
   \label{Fig.3}
\end{figure}

\subsection{Real-time Simulations of Plasmon Excitation of Au$_{20}$-CO}
\noindent
The processes of plasmon relaxation in the Au$_{20}$-CO system are summarized in Figure~\ref{Fig.3}. The NAMD simulations allow us to delineate different processes during plasmon relaxation, as shown schematically in Figure~\ref{Fig.3}(a). To analyze the energy evolution during the relaxation process, a time-energy 2D map is constructed by calculating the probability of finding the system at a given energy level and time. Specifically, the probability is defined as
\begin{equation}
    P(E,t)=\frac{N_{E}(t)}{N_{tot}},
\end{equation}
where $N_{E}(t)$ and $N_{tot}$ stand for the number of trajectories with excitation energy $E$ and the total number of trajectories, respectively. Such representation is similar to that often used in experiments,\cite{chan2011observing} and thus provides a direct comparison between simulation and experimental data. Figure~\ref{Fig.3}(b) shows the time-energy 2D map for Au$_{20}$-CO. The non-monoexponential decay of excitation energy and a faster relaxation compared to bare Au$_{20}$ (Figure~S5) indicate that multiple vibrational modes, in particular CO stretching, participate in the relaxation process of plasmon excitation. Taking the average excitation energy ($\overline{E}(t) =\int E P(E,t) dE$) at each time step, the lifetime of energy relaxation can be obtained by fitting the data to an exponential decay function 
\begin{equation}
    \overline{E}(t) = Ae^{-t/\tau} + B,
\end{equation}
where $\tau$ is the energy relaxation time. $A$ and $B$ are determined according to the conditions: $A+B= \overline{E}(0)$, $B=\overline{E_{1}}(\infty)$, and $E_{1}$ stands for the energy of the lowest excited state ($S_{1}$). From Figure~\ref{Fig.3}(c), the lifetime of energy relaxation process in Au$_{20}$-CO is estimated to be $\sim$ 1.0~ps, which is significantly faster than that of bare Au$_{20}$ ($\sim$2.7~ps). Here, non-adiabatic coupling between ground and excited states is not considered. Thus, the system is allowed to relax to $S_{1}$ only. This result indicates that hybridization of the MOs of CO molecule and Au$_{20}$ accelerates the energy relaxation process by offering an additional pathway for plasmon energy decay. It is postulated that the CO stretching mode participates in the relaxation process via the CT excited states, which accelerates the energy relaxation. 

\begin{figure}[htb]
   \centering
   \includegraphics[width=0.95\textwidth]{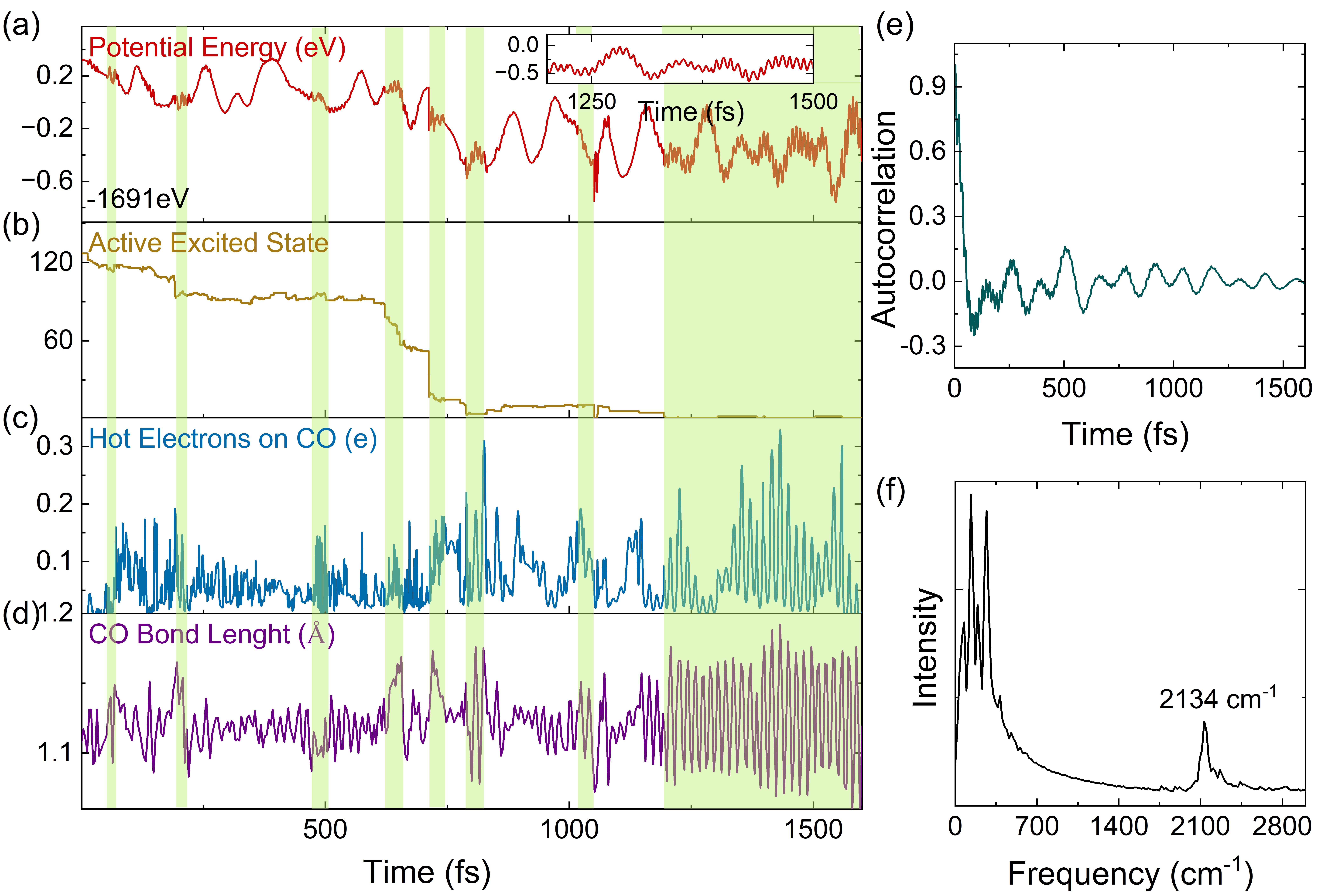}
   \caption{
   An analysis of a characteristic trajectory. From top to bottom: (a) potential energy with time, (b) active excited state, (c) transition charges on CO, and (d) variation of bond length of CO. (e) Autocorrelation function of velocity. (f) Phonon spectra obtained by Fourier transform of (e). Inset in (a) shows an enlarged view of potential energy to highlight the high frequency oscillations. 
   }
   \label{Fig.4}
\end{figure}

Apart from ensemble average of all the trajectories in the dynamics, each trajectory in our simulations give insights into different possible reaction pathways. Here, we analyze a typical trajectory to shed light on the details of plasmon relaxation. Figure~\ref{Fig.4}(a-d) shows the evolution of the potential energy, excited state occupations, number of HEs on the CO molecule, and the bond length of CO. The situation when the potential energy curve shows vibrations with high frequency and small amplitude is highlighted with green shadow. The appearance of this vibration is found to be associated with non-adiabatic hops between excited states, as shown in Figure~\ref{Fig.4}(b). The hops between excited states is accompanied with electron transfer to CO molecule (Figure~\ref{Fig.4}(c)) and also an elongation of C-O bond (Figure~\ref{Fig.4}(d)). Such a strong concurrent behavior across all these quantities, especially the coincidental trend between HEs on CO and vibrational amplitude of C-O bond length at the moment of hopping, directly reveals that the HEs transfer is concomitant to the vibrational excitation of CO. From Figure~\ref{Fig.4}(c), it can be seen that HEs transfer back and forth between Au$_{20}$ and CO repeatedly, showing energy exchange between the CO vibrational mode and the HEs. To obtain further insights into electron-vibration scattering, we identify the vibrational modes that are coupled to the electronic subsystem. For this purpose, the auto-correlation function of the velocity is calculated and plotted in Figure~\ref{Fig.4}(e). The dominant vibrational modes contributing to the electron-vibration coupling can be identified from the Fourier transform of the auto-correlation function. As shown in Figure~\ref{Fig.4}(f), it can be clearly seen that C-O stretching mode with a frequency of 2134~cm$^{-1}$ is excited during the relaxation process. It is noted that the chemisorption on Au$_{20}$ leads to a red shift of stretching frequency with respect to 2395.6~cm$^{-1}$ in gas-phase calculated by DFTB method. On the other hand, the low-frequency vibrational modes in the range of 0–400~cm$^{-1}$ are attributed to the Au$_{20}$ vibrations and the vibrations induced by the Au$_{20}$-CO coupling.

\begin{figure}[!htb]
   \centering
   \includegraphics[width=0.95\textwidth]{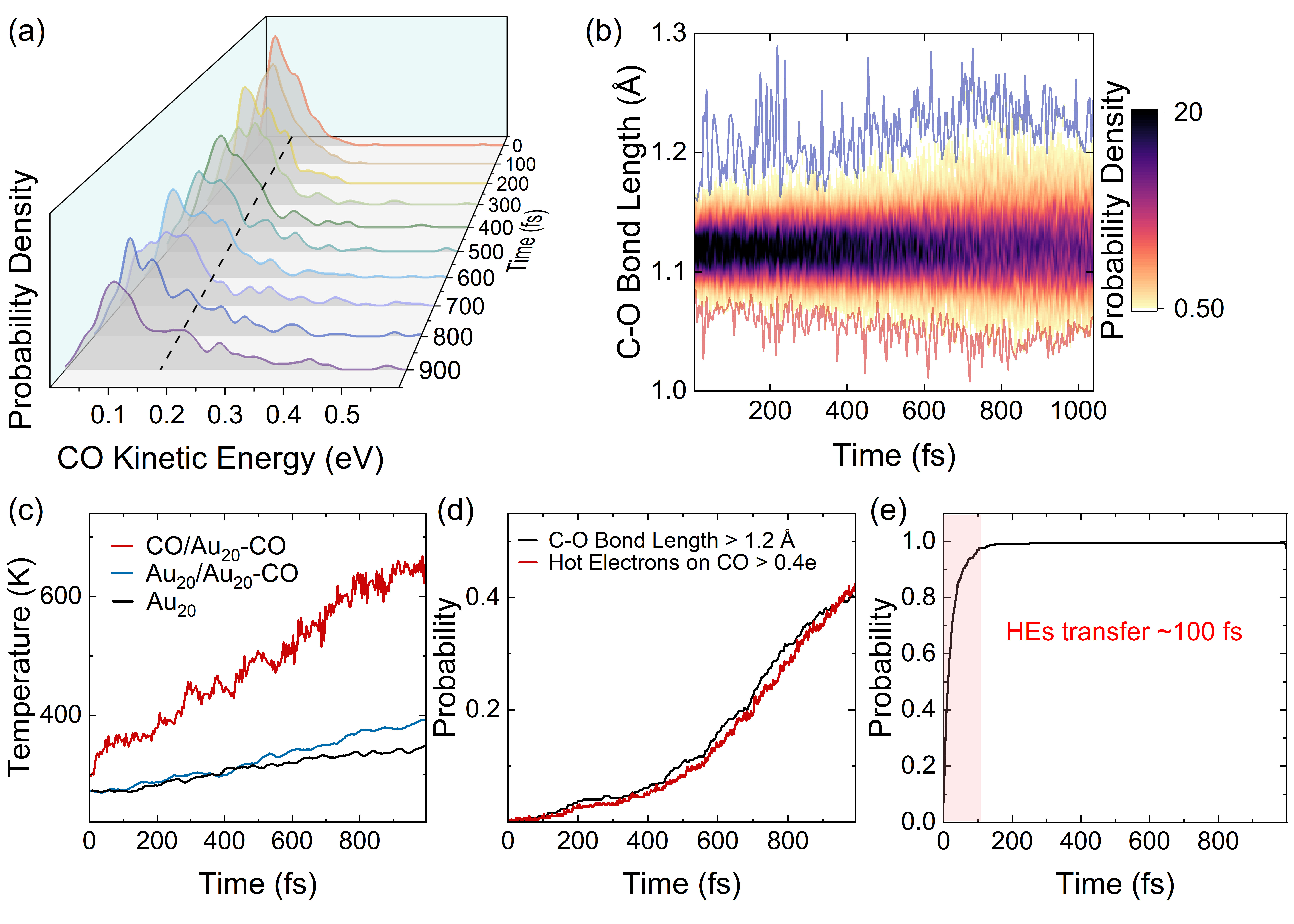}
   \caption{An analysis of the ensemble of trajectories. (a) Distributions of CO kinetic energy with time. (b) Distributions of the C-O bond length with time. (c) Evolution of the local temperature with time. (d) The net probability of the C-O bond length being larger than 1.2~{\AA} and the HEs on CO being greater than 0.4$e$  with time. (e) Accumulative probability of the transition charges on CO greater than 0.1$e$. All results are the statistical results averaged over 300 trajectories.
   \iffalse
   \yz{YZ: reorder the labels in sequential order: top a-b, bottom c-e; And b) is discussed at the end, it should be e)}.\fi  }
   \label{Fig.5}
\end{figure}

We next calculate the ensemble averages of kinetic energies, HE distributions, to obtain experimental relevant observables and estimate the efficiency of plasmon-induced CO activation based on these statistical results. An increase in the kinetic energy of CO is highlighted by the waterfall plots in Figure~\ref{Fig.5}(a). The average kinetic energy of CO, as marked by the dashed line, increases from 0.08 to 0.16~eV over the 1~ps relaxation timescale. Figure~\ref{Fig.5}(b) shows the evolution of the bond length of CO, where the red and blue curves represent the shortest and longest bond lengths at each time step, respectively. It can be clearly observed that the bond length of CO oscillates around 1.12~{\AA}, and its oscillation amplitude gradually increases. We then estimate the ionic temperature of CO and Au$_{20}$ according to the equipartition theorem ($E_{kin}=\frac{3}{2}Nk_{B} T$, $k_{B}$ is the Boltzmann constant, $E_{kin}$ is the kinetic energy). As shown in Figure~\ref{Fig.5}(c), the temperature increase of CO is significantly faster than that of Au$_{20}$, and the final temperature of CO is 4 times higher than that of Au$_{20}$ after the 1~ps simulation. This steep temperature gradient strongly indicates that activation of CO originates from HE transfer-induced vibrational excitations rather than thermal effect. According to Figure~\ref{Fig.5}(b), we define CO to be activated if the amplitude of CO vibrational motion is larger than 1.2~{\AA}, and the corresponding probability of CO activation is shown in Figure~\ref{Fig.5}(d). The evolution of HEs on CO shows a similar trend with the increase of CO bond length, and the probability of activation reaches $\sim$40\% at 1~ps, demonstrating that the HEs transfer can effectively activate CO by exciting the C-O stretching mode.

Finally, we estimate the timescale of HEs transfer from Au$_{20}$ to CO by calculating the accumulated probability that HEs on CO is greater than 0.1$e$ at each time step. However, the HEs on CO oscillate with time, indicating back-and-forth CT between Au$_{20}$ and CO. In order to analyze the accumulated probability, we think that for each trajectory, once the HEs on CO is greater than 0.1$e$ at a certain time, the HEs on CO at all subsequent time steps is set to 0.1~$e$ in the statistics. Here, the reason we choose 0.1$e$ as the target value for determining whether HEs transfer occurs is that the initial plasmon excitation has $\sim$ 0.1$e$ charge transfer from Au$_{20}$ to CO. As shown in Figure~\ref{Fig.5}(d) and Figure S4, initially, a majority of HEs are distributed on Au$_{20}$, and the probability of direct HEs transfer is only $\sim$7\%. However, HEs rapidly transfer to CO (0.1$e$) within 100~fs. Such a fast process is also consistent with previous experimental investigations.~\cite{bauer2006hot} Importantly, the energy relaxation time ($\sim$1~ps) is significantly longer than the HE transfer timescale ($\sim$100~fs), indicating that HE transfer occurs before electronic energy relaxation completes. Therefore, it is expected that the plasmon excitation in Au$_{20}$ can effectively activate CO adsorbate and mediate subsequent chemical reactions before the nonradiative decay of HEs.

\subsection{Conclusion}  
The plasmon-induced bond activation of CO adsorbed on Au$_{20}$ cluster triggered by the HEs transfer has been investigated by a non-adiabatic molecular dynamics simulation combining LR-TDDFTB theory and TSH method. The simulations naturally treat the entire dynamical processes and multiple interactions in plasmon-mediated chemistry on an equal footing, including the plasmon excitation, HEs relaxation, direct$\slash$indirect HE transfer, and the activation of CO vibration mode induced by HE transfers. These dynamical processes compete with the vibrations scattering process inside Au$_{20}$ that results in local heating of the system. Our simulations provide a comprehensive picture of plasmon-induced chemistry in atomic details.

The simulations demonstrate that CT states are crucial in plasmon-induced CO activation. The direct and indirect HEs transfers are generated by excitation of the CT states, leading to the activation of CO stretching mode, which is also an essential feature of the plasmon energy relaxation process. This activation of CO vibrational mode has been observed with high efficiency ($\sim40\%$). Importantly, the HEs transfer is faster than the conventional scattering process of the Au$_{20}$. The simulations show that HEs transfer from Au$_{20}$ to CO completes within ${\sim}$100~fs, while the energy relaxation occurs on a ${\sim}$1~ps timescale. 

The present simulation employs the LR-TDDFTB-based TSH method to simulate non-adiabatic molecular dynamics following the plasmon excitation directly. At each time step, 160 excited states are calculated based on the LR-TDDFTB method, pushing the limits of currently feasible theoretical methods. 
Nevertheless, our direct atomistic simulations outline the dynamics of the plasmon-mediated chemical transformations in terms of the evolution of the potential energy with the reaction coordinates. This enables a straightforward demonstration of the energy relaxation and HE transfer during the reaction and elucidates the details of the reaction pathways. Our numerical simulations give clear dynamical insights into CO bond activation on Au clusters. We believe the study reported in this manuscript paves the way for simulating and understanding plasmon-mediated photochemistry.

\begin{acknowledgement}
C.Y.Y. acknowledges the support from the NSFC (Grant Nos. 22073007 and 12088101), the Guangdong Shenzhen Joint Key Fund (Grant No.2019B1515120045), the Shenzhen Basic Research Fund (Grant No. JCYJ20190813164805689), the Sino-German mobility program (Grant No. M-0215), and Hong Kong Quantum AI Lab. Y.Z. and S.T. acknowledge support from LDRD program of Los Alamos National Laboratory (Grant No. 20220527ECR). S.T. acknowledges support of the Humboldt Research Award (Germany) and the Center for Integrated Nanotechnology (CINT) at Los Alamos National Laboratory (LANL), a U.S. Department of Energy and Office of Basic Energy Sciences User Facility. LANL is operated by Triad National Security, LLC, for the National Nuclear Security Administration of U.S. Department of Energy (contract no. 89233218CNA000001). 
\end{acknowledgement}

\begin{suppinfo}

Additional figures and tables. 

\end{suppinfo}

\bibliography{CO_activation}

%\section{Figure}
%\yz{Figures 1 and 2 can be combined together (only need to do it in the final stage)}.

\end{document}